\documentclass[useAMS,usenatbib,intlimits,centertags,usegraphicx]{mn2e}
\usepackage{amsmath,amssymb}
\usepackage[latin1]{inputenc}
\usepackage{graphicx}
\usepackage{citesort}
\usepackage{booktabs}
\usepackage{color}

\newcommand{\ud}{\mathrm{d}}
\newcommand{\be}{\begin{equation}}
\newcommand{\ee}{\end{equation}}
\newcommand{\pt}{\partial}

\newcommand{\omg}{\Omega}

\newcommand{\e}{{\rm e}}

\begin{document}

\title[Differentially Rotating Neutron Stars]{On the Solution Space of Differentially Rotating Neutron Stars in General Relativity}


\author[Ansorg, Gondek-Rosi\'{n}ska \& Villain]
 {Marcus Ansorg$^1$, 
  Dorota Gondek-Rosi\'{n}ska$^{2,3,4}$ and Lo\"\i{}c Villain$^{5,4}$\\
  $^1$Max-Planck-Institut f\"ur Gra\-vi\-ta\-ti\-ons\-phy\-sik, 
                Albert-Einstein-Institut, Am M\"uhlenberg 1, 14476 Golm, Germany \\
  $^2$ Institute of Astronomy, University of Zielona G\'ora, Lubuska 2, 65-265, Zielona G\'ora, Poland\\
  $^3$  Nicolaus Copernicus Astronomical Center, PAN, Bartycka 18, 00-716 Warsaw, Poland\\
  $^4$ LUTH, CNRS, Observatoire de Paris, F-92195 Meudon Cedex, France\\
  $^5$ DFA, Universitat d'Alacant, Ap. Correus 99, 03080 Alacant, Spain
  }

\maketitle

\allowdisplaybreaks

\begin{abstract}

A highly accurate, multi-domain spectral code is used in order to construct sequences of general relativistic,
differentially rotating neutron stars in axisymmetry and stationarity. For bodies with a spheroidal topology and a homogeneous or an $N=1$ polytropic equation of state, we investigate the solution space corresponding to broad ranges of degree of differential rotation and stellar  densities. In particular, starting from static and spherical configurations, we analyse the changes of the corresponding surface shapes as the rate of rotation is increased. For a sufficiently weak degree of differential rotation, the sequences terminate at a mass-shedding limit, while for moderate and strong rates of differential rotation, they exhibit a continuous parametric transition to a regime of toroidal fluid bodies. In this article, we concentrate on the appearance of this transition, analyse in detail its occurrence and show its relevance for the calculation of astrophysical sequences. Moreover, we find that the solution space contains various types of spheroidal configurations, which were not considered in previous work, mainly due to numerical limitations.
\end{abstract}

\section{Introduction}

Rotating relativistic stars are among the most promising sources of gravitational waves (see Sathyaprakash and Schutz 2009 for a review), especially when they are newly born, be it as proto-neutron stars resulting from a gravitational core collapse or as compact remnants of neutron stars binary mergers. Yet, their study is a complicated task, involving the solution of Einstein's equations in a dynamical regime while, at the same time, many microphysical phenomena need to be taken into account. Therefore, one of the first compulsory steps in their modelling is the analysis of axisymmetric and stationary configurations of rotating self-gravitating perfect fluids. However, even when a constant rotation profile is assumed, the solution space of this academic problem is known to possess a complex structure (see
Ansorg {\it et al.} 2004; Meinel {\it et al.} 2008), which implies that robust and precise numerical codes are needed before one can proceed with the astrophysical investigation. 

A well-studied example is the solution space of
homogeneous, uniformly rotating bodies to which, in Newtonian gravity,
the famous Maclaurin spheroids belong. For given mass density, this
family depends, but for a scaling factor, on only one parameter, for
which reason we may refer to it as the {\em Maclaurin
  sequence}. However, the full solution space is much richer as it
contains infinitely many Newtonian sequences that branch off from the
Maclaurin sequence (see Ansorg {\it et al.} \cite{AKM03b}). Taking
these configurations into the realm of general relativity, one
recognizes the entire complexity of the solution space which turns out
to consist of infinitely many disjoint solution classes (see Ansorg
{\it et al.} \cite{Ansorgetal04}).  Within these classes (which are
two-dimensional subsets of the solution space), specific sections of
the above Newtonian sequences form particular limiting curves.

 One consequence of this class
 structure is that any continuous sequence of relativistic solutions starting at a static and
 spherical body and running through the associated {\em spheroidal class},
 is not connected to the regime of configurations with toroidal
 shape. Instead, such a sequence eventually leads either to mass shedding
 limit\footnote{A mass-shedding limit is given if a fluid particle at the
 surface of the body moves with the same angular velocity as a test particle
 at that spatial point. The corresponding geometrical shape of the fluid body
 possesses a cusp there.} or to a limit of infinite central pressure. 

In a slightly more realistic model of a star, which one obtains by
taking polytropic matter, a picture similar to the one depicted above
for homogeneous bodies emerges (Meinel {\it et al.}
\cite{Meineletal08}). As one considers polytropes with increasing
polytropic index $N$ (with $N=0$ describing the homogeneous bodies),
one finds that the different classes drift apart from one another,
implying an even clearer separation between them.

However, apart from the equation of state (EOS),
actual stars differ from the model of simple homogeneous rigidly
rotating fluids through their presumable differential rotation, which is expected to occur in proto-neutron stars born from a gravitational collapse or from the merger of binary neutron stars. Hence,
a natural question that arises is that of the influence of the
rotation profile on the main characteristics of a stationary,
relativistic star, such as its maximal mass and
  angular momentum, but also its actual geometrical surface shape. The evolution of the latter one 
  possibly leads to various types of instabilities and corresponding
  signals (e.g. electromagnetic or gravitational waves,
  neutrinos). Due to its importance for estimating the delay between a neutron stars binary merger and the collapse to a black hole, this issue was already the topic of various studies, be
it for cold or warm equations of state (see Stergioulas \cite{S} for a
review on rotating stars in relativity). However, in the case of
differentially rotating stars, most of the work done up to now seemed
to suffer from numerical limitations when the rotation profile started
to be too sharp. Moreover, none of those studies really dealt with the
changes undergone by the solution space due to the rotation law.

In this article, we focus on the influence of differential
rotation on the parameter space of rotating relativistic stars\footnote{We postpone the discussion of astrophysically relevant quantities (such as the mass, the angular momentum) to some other article.}.
In particular, we only consider star-like bodies, i.e.~configurations with spheroidal topology
(simply connected ones), and use two basic EOSs (homogeneous and polytropes with $N=1$) as well as one specific type of rotation law. The latter one
is the most classical in general relativity, introduced by Komatsu
{\it et al.} \cite{KEH}, in which a parameter $A$ appears that
measures the degree by which the rotation is differential. This allows
us to go from rigid rotation to configurations with a strong gradient
in the rotation profile. As will be described below, in this situation
some critical values of $A$ arise, at which a merging of typical
classes is exhibited. This result shows that the introduction of several {\em disjoint}
classes, initially done in order to characterize the solution space
corresponding to {\em uniform rotation}, only makes sense for
sufficiently weak differential rotation\footnote{Hereby, these
critical values of $A$ depend sensitively on the EOS being
chosen. Below we demonstrate that for $N=1$ polytropes the class
structure can still be introduced for sufficiently weak differential
rotation. In contrast, for homogeneous matter the class structure
ceases to exist in a strict sense for {\em any} rate of differential
rotation.}. 


The paper is organized as follows. First, we clarify the distinction
between the so-called {\em spheroidal class} and the {\em toroidal
  class} of rotating stars, a notation that should not be confused
with the star's topology, see discussion below. Then, in Section 3, we
introduce the Lewis-Papapetrou line element that describes stationary
and axisymmetric spacetimes. At the end of that Section we discuss the
Komatsu law of differential rotation, to be used in this paper, as
well as the resulting general relativistic field equations. In Section
4 we briefly describe particular features of our pseudo-spectral code,
used to obtain highly accurate numerical solutions and present
numerical results. We introduce four specific types of sequences of
differentially rotating relativistic stars, which describe the
solution space in question independently of the EOS being chosen.
Finally Section 5 provides a summary and a discussion of our results.

\section{Spheroidal and toroidal classes}

In the following we shall concentrate on two specific classes, which we call {\em spheroidal} and {\em toroidal}. For rigidly rotating stars, the first one has already been mentioned in the Introduction and contains continuous parametric sequences that start at a static and spherical body. The shapes of these bodies possess spheroidal topology throughout the entire class. Any such sequence eventually leads either to a mass-shedding limit or to a limit of infinite central pressure. For this reason it makes sense to identify extremal values for certain physical quantities, such as the maximal baryon mass, within the spheroidal class.

For the introduction of the {\em toroidal} class we consider the so-called {\em Dyson ring sequence}, which is the Newtonian sequence of rigidly rotating homogeneous bodies branching off from the Maclaurin sequence at the branch point, at which the Maclaurin spheroids become secularly unstable with respect to the first axisymmetric perturbation (Chandrasekhar \cite{chand67}, Bardeen \cite{B71}). One branch of the Dyson ring sequence terminates at a mass-shedding limit, while another one  leads to the regime of figures with toroidal topology and finally to arbitrarily thin rings\footnote{We call a ring {\em thin} if the ratio of inner to outer circumferential equatorial radius is close to 1.}. Taking the Dyson ring solutions into the relativistic regime and choosing again homogeneous matter, we find an associated {\em toroidal} class of solutions which is disjoint with respect to the {\em spheroidal} class. In contrast to the spheroidal class, it possesses continuous sequences of relativistic solutions leading to infinitely thin tori. Moreover, the solutions of the toroidal class do not possess maximal possible values of the baryon and gravitational masses. Instead, those quantities increase to arbitrarily high values as the relativistic tori become thinner and thinner (see Ansorg {\em et al.} \cite{AKM03c} and Fischer {\em et al.} \cite{Fischeretal05}). 

As explained in the Introduction, the solution space of rigidly rotating $N=1$ polytropes is quite similar, except that even in the Newtonian limit there is already a clear gap between the spheroidal and toroidal sequences. This gap grows further as polytropes with increasing index $N$ are considered. 

As an important remark, note that in the above definition of spheroidal and toroidal classes, the existence of continuous transitions, either to spheres or to infinitely thin tori, is considered. This does not determine {\em a priori} the geometrical topology of the configurations which form the sequence. This is to say that a body with spheroidal topology (i.e.~which is simply connected and has no hole) may in fact belong to the toroidal class, provided that a continuous parameter transition to the regime of infinitely thin tori can be identified. 

In this article we focus, for the purpose of simplicity, on stars with
spheroidal topology and find that the situation becomes more
complicated for differentially rotating bodies. In particular, a merging of the spheroidal and toroidal classes is exhibited, meaning that from some critical rate of differential rotation on the classification does not make sense\footnote{Note that the numerical code needs to be specifically adapted to the treatment of configurations with rotation rates close to a critical value, in order to avoid non-convergence or random numerical jumps between different types of solutions.}. However, as we shall see below, a discussion of the corresponding solution space is possible in terms of four different {\em types} (A,B,C and D) of one-dimensional parameter sequences. 

\section{Differentially Rotating Spheroidal Stars in General Relativity}

\subsection{Line element}
Using Lewis-Papapetrou coordinates $(\varrho, z, \varphi, t)$, the line element for a stationary, axially
symmetric spacetime corresponding to a rotating perfect 
fluid configuration can be written in the form (see e.g.~Stephani {\em et al.} \cite{Stephani_etal})
\be\label{metric}
        \ud s^2=\e^{2\mu}(\ud\varrho^2+\ud z^2)+W^2\e^{-2\nu}(\ud\varphi-\omega\,\ud t)^2-
       \e^{2\nu}\,\ud t^2,
\ee
where the metric functions $\mu$, $W$, $\nu$ and $\omega$ are functions of
$\varrho$ and $z$ alone. Note that
\be\label{e:W}
        W = 0 \qquad \mbox{on the rotation axis $\varrho=0$.}
\ee
The coordinates $(\varrho,z,\varphi,t)$ are
uniquely defined by the requirement that the metric coefficients and their first derivatives 
be continuous everywhere, in particular at the surface of the rotating body. 

\subsection{Equation of state and Rotation law}

For a barotropic perfect fluid, the energy-momentum tensor reads
\be\label{Tik}
        T^{ik}=[\epsilon(p)+p]u^iu^k+pg^{ik}\,,
\ee
where $u^i$ is the fluid's 4-velocity. The total mass-energy density $\epsilon$ as a function of the pressure $p$ is given via an equation of state. 

In this paper we consider neutron stars described by polytropic or homogeneous EOSs.
The polytropic equation of state is given by  
\be
        \epsilon\,(p)=p+\sqrt{p/K}
\ee
where $K$ is the polytropic constant. The polytropic index $N$ is chosen to be unity as this value fits well most of the realistic EOSs for neutron stars (see Lattimer \& Prakash \cite{LattimerP2001}). Since this is no longer true for strange stars (e.g. Limousin {\it et al.} \cite{LimousinGG2005}), whose properties can be much better reproduced using a homogeneous model (see Gondek-Rosi\'nska {\it et al.} \cite{GondekGH2003}, Amsterdamski {\it et al. }\cite{Amster2002}), we also use such an EOS with
\be
        \epsilon(p) = \epsilon_0 = \mbox{constant.}
\ee.

The stationary and circular motion of the fluid particles implies a 4-velocity of the form:
\be\label{four}
        u^i=\e^{-V}(\xi^i+\Omega\eta^i)
\ee
where $\xi^i$ and $\eta^i$ are the Killing vectors with respect to stationarity and axisymmetry respectively. The coefficient $\Omega=u^t/u^\varphi$ denotes the angular velocity of the fluid and depends on $\varrho$ and $z$. The integrability conditions of the Einstein equations, i.e. the Euler equations
\be
   T^{ik}_{\;\;\;\;;k} = 0
\ee
restrict the possible form that the angular velocity can take. They imply
\be
        -\frac{\ud p}{\epsilon+p} = \ud V + u^t u_\varphi \ud\Omega
\ee
from which it follows that the product $u^t u_\varphi$ is a function depending on $\Omega$ only,
\be\label{Func_F}
        u^t u_\varphi = F(\Omega),
\ee
and that the specific enthalpy 
\be\label{e:enthalpy}
        h=\frac{\epsilon+p}{\epsilon_B}
\ee
(where $\epsilon_B$ denotes the {\em baryon} mass density) with
\be \label{e:d_enthalpy}
   \frac{\ud h}{h}=\frac{\ud p}{\epsilon+p}
\ee
is related to the metric quantities as follows:
\be\label{e:int_enthalpy}
        h = h_0\exp\left(V_0-V-\int_{\omg_c}^\omg F(\tilde{\omg}) d\tilde{\omg}\right).
\ee
Here, the constant $\omg_c$ denotes the central angular velocity, while $V_0$, i.e.~the value of the function $V$ at the fluid's north pole ($\varrho=0,\; z=r_\mathrm{p}$), is a relativistic parameter characterizing the rotating fluid body. It is related to the relative redshift $z_0$ of photons emitted from the north pole and received at infinity via
\[
        z_0=\e^{-V_0} - 1.
\]
Note that from $u^i u_i = -1$ we have
\be
        V = \frac{1}{2}\log\left[\e^{2\nu} - W^2 \e^{-2\nu}(\Omega-\omega)^2\right]
\ee
and
\be
        u^t u_\varphi = \frac{W^2 (\Omega-\omega)}{\e^{4\nu} - W^2 (\Omega-\omega)^2}.
\ee
Hence, by virtue of (\ref{Func_F}) we obtain $\Omega$ in terms of the metric coefficients as the solution of the implicit equation
\be\label{e:Det_Omega}
        F(\Omega) = \frac{W^2 (\Omega-\omega)}{\e^{4\nu} - W^2 (\Omega-\omega)^2}.
\ee
It follows from (\ref{e:W}) and (\ref{e:Det_Omega}) that (i) $F(\omg_c)=0$ and (ii) $\omg=\omg_c$ on the rotation axis.

The enthalpy $h_0$ at vanishing pressure takes, for ordinary matter (such as the ones considered here), the value $h_0=1$.\footnote{An exception is so called strange quark matter, see e.g.~Gourgoulhon {\it et al.} \cite{Gourg1999}.}

The function $F$ can be chosen freely, being only subject to the condition $F(\omg_c)=0$.
In this paper we use the simplest form which goes beyond uniform rotation, i.e. we take a linear ansatz
\be\label{e:lawdif}
 F(\omg)\,=\,A^2\,(\omg_c\,-\,\omg)\,,
\ee
with the constant $A$ describing the rate of differential rotation. This particular rotation law, originally introduced by Komatsu {\it et al.} \cite{KEH}, has been considered by various authors (including Cook {\it et al.}
\cite{CST}, Bonazzola {\it et al.} \cite{BGSM}, Goussard {\it et al.}
\cite{GHZ}, Baumgarte {\it et al.} \cite{Bau00}, Lyford {\it et al.} \cite{LBS}, Morrison {\it et al.} \cite{MBS}, and Villain {\it et al.} \cite{VPCG}). Since the typical scale of length of the problem is the radius of the star and
since the rotation profile becomes uniform in the limit $A \to \infty$,
we shall not use $A$ directly, but following Baumgarte {\it et al.} \cite{Bau00}, we parameterize the sequences and the degree of differential
rotation using the parameter
\be\label{e:tilde_A}
        \tilde{A}=\hat{A}^{-1}=r_\mathrm{e}/A\,,
\ee
where $r_\mathrm{e}$ is the star's equatorial coordinate radius\footnote{Notice that we
shall not use the $\hat{A}$ variable, but mention it so as to make a comparison with articles from other authors easier.}. Without entering more into the detail of the law (\ref{e:lawdif}) [see
Stergioulas \cite{S} for further information], we note that it 
verifies Rayleigh's criterion for stability against axisymmetric perturbations
(Komatsu {\it et al.} \cite{KEH}) and that, in the Newtonian limit, it reduces
to 
\be
\label{omelaw}
\omg\,=\,\frac{\omg_c A^2}{ A^2 + \varrho^2}\,,
\ee
which can be shown to be in reasonable agreement with configurations of newly
born neutron stars resulting from core collapse (see Section 4 of Villain {\it
  et al.} \cite{VPCG}) or from dynamical merger simulations (see Lyford {\it
  et al.} \cite{LBS}).\\

\subsection{Einstein equations and the free boundary value problem}

The set of equations, following from Einstein's field equations $R_{ik}-\frac{1}{2}Rg_{ik}=8\pi T_{ik}$
for the metric in the form (\ref{metric}), with (\ref{Tik}) and (\ref{four}), can be written as follows (see e.g.~Bardeen \& Wagoner \cite{BW71}):
\begin{subequations}\label{field_eqsa}
 \begin{align}
   \nonumber
        & \nabla\cdot(B\nabla\nu)  - \frac{1}{2}\varrho^2B^3\e^{-4\nu} (\nabla \omega)^2  = \\
        & \hspace*{2cm} 4\pi \e^{2\mu}B\left[(\epsilon+p)\frac{1+v^2}{1-v^2} + 2p \right],
 \\[3mm]
  & \nabla \cdot (\varrho^2 B^3 \e^{-4\nu} \nabla \omega) = 
    -16\pi \varrho B^2 \e^{2\mu-2\nu}\, (\epsilon+p)\, \frac{v}{1-v^2},
\\[3mm]
   & \nabla \cdot (\varrho\nabla B) = 16\pi \varrho B \e^{2\mu} p \label{eqB}
\\[3mm]
 \nonumber & \mu_{,\varrho\varrho}+\mu_{,\zeta\zeta} - \frac{1}{\varrho}\nu_{,\varrho} 
 +\nabla\nu\left(\nabla\nu-B^{-1}\nabla B\right) = \\
  & \hspace*{2cm}  \frac{1}{4}\varrho^2 B^2\e^{-4\nu}(\nabla\omega)^2
    -4\pi \e^{2\mu}(\epsilon+p),
\intertext{with [cf. (\ref{e:W})]}
 &B := W/\varrho \quad 
    \text{and} \quad v := W \e^{-2\nu} (\Omega-\omega). \label{vrot}
 \end{align}
\end{subequations}
Here the operator $\nabla$ has the same meaning as in a Euclidean three-space in
which $\varrho$, $z$ and $\varphi$ are cylindrical coordinates.

The equations (\ref{e:d_enthalpy} - \ref{e:lawdif}) allow us to express $p, \epsilon$ and $\Omega$ and hence all source terms in (\ref{field_eqsa}) with respect to the metric potentials and the three constants $\Omega_c, A$ and $V_0$. Thus we obtain a complete set of elliptic equations for the four metric coefficients $\mu$, $B$, $\nu$ and $\omega$. 

Note that the fluid's surface shape is unknown and must be determined in the context of the numerical solution scheme. Its definition is given through vanishing pressure (i.e. through $h=h_0$), which leads by virtue of (\ref{e:int_enthalpy}) to the surface condition:
\be
        V = V_0 - \int_{\omg_c}^\omg F(\tilde{\omg}) d\tilde{\omg}
\ee
The asymptotic conditions at infinity:
\be
        \mu=\nu=\omega=0, \qquad B = 1,
\ee
complete the free boundary value problem to  be solved. For a given EOS, the corresponding solutions depend on the three parameters $\Omega_c, A$ and $V_0$.

\begin{figure}
        \vspace*{-1.7cm}\hspace*{-2.6cm}\includegraphics[scale=0.9]{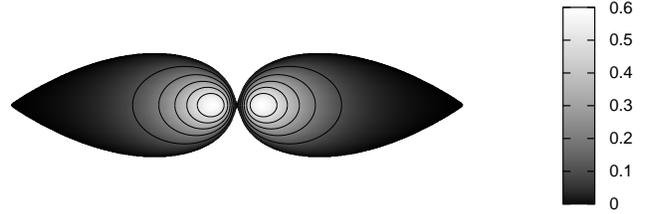}\vspace*{-2.7cm}
        \caption{Example of a differentially rotating polytropic body (with polytropic index $N=1$) close to the transition of spheroidal to toroidal topology. The parameters of this configurations are given by $\tilde{A}=0.9, \log h_\mathrm{max}=0.6$ and $r_\mathrm{p}/r_\mathrm{e}=0.005$. In the corresponding diagram in fig.~\ref{fig3} this body belongs to the sequence which terminates in a mass-shedding limit ($\tilde{\beta}=0$) at $r_\mathrm{p}/r_\mathrm{e}\approx0.25$. The plot shows isocontours of $\log h$ 
in a meridional cross section. The strongly pronounced cusp at the equatorial edge indicates that this configuration rotates almost at the mass-shedding limit.}
        \label{fig1}
\end{figure}

\section{Numerical results}

\subsection{Numerical scheme and parameter prescriptions}

We use a pseudo-spectral double-domain method to compute highly accurate numerical solutions corresponding to rotating relativistic stars. This technique has been described in detail in Ansorg {\it et al.} \cite{Ansorgetal02, AKM03a}, see also Meinel {\it et al.} \cite{Meineletal08}. Here we develop the scheme further to include the calculation of differentially rotating stars. In particular, strongly pinched objects can be computed (for an example see fig.~\ref{fig1}).

As a special feature, the code permits the prescription of three independent
parameters which may be chosen for convenience. For example, it is possible to
prescribe the parameter triple $(h_\mathrm{max}, r_\mathrm{p}/r_\mathrm{e},
\tilde{A})$ where $h_\mathrm{max}$ denotes the maximal enthalpy of the
star. Note that this maximal value need not be assumed in the coordinate
origin $\varrho=z=0$ but can, for stars sufficiently close to the toroidal
regime, be located at $(\varrho_h,z=0)$ with $\varrho_h>0$, see
e.g.~fig.~\ref{fig1}. The dotted line on each diagram of figures \ref{fig3} and
\ref{fig5} separates configurations with the maximal enthalpy located in the
stellar center (on the right side) from those having $h_\mathrm{max}$  outside
the coordinate origin. This curve is defined by the vanishing of the
enthalpy's second derivative with respect to $\varrho$ at the origin,
i.e. through $(\ud^2h/\ud\varrho^2)_c = 0$. We determine such a configuration
through the prescription of the triples\footnote{Note that for triples of
  prescribed parameters, which do not contain the rotation law parameter
  $\tilde{A}$, the corresponding value of this quantity is determined in the
  numerical calculation. Likewise the parameters $V_0$ and $\Omega_c$ are part
  of the computational result.} $(h_\mathrm{max}, r_\mathrm{p}/r_\mathrm{e}, (\ud^2h/\ud\varrho^2)_c)$ or $(h_\mathrm{max}, \tilde{A}, (\ud^2h/\ud\varrho^2)_c)$.

Another possible parameter prescription is the triple $(h_\mathrm{max}, r_\mathrm{p}/r_\mathrm{e}, \beta)$ where the shedding parameter $\beta$ is defined as in Ansorg {\it et al.} \cite{AKM03a},
\be\label{beta}
        \beta=\left.-\frac{r_\mathrm{e}^2}{r_\mathrm{p}^2}\;\frac{\ud (z_b^2)}{\ud (\varrho^2)}\right|_{\varrho=r_\mathrm{e}}
\ee
with $z=z_b(\varrho)$ describing the star's surface. This parameter is chosen such that $\beta=0$ in the mass-shedding limit and $\beta=1$ for (Maclaurin) spheroids. If a sequence exhibits the transition to the toroidal regime (i.e. $r_\mathrm{p}/r_\mathrm{e} \to 0$), $\beta$ tends to infinity as this transition is encountered. For this reason we prefer to consider the rescaled shedding parameter
\be\label{tilde_beta}
        \tilde{\beta} = \frac{\beta}{1+\beta}
\ee
which tends to 0 in the mass-shedding limit, to $\frac{1}{2}$ for the Maclaurin spheroids as well as in the spherical non-rotating limit and to 1 at the transition point to the toroidal regime.

For the exact calculation of the critical rotation law parameter $\tilde{A}_\mathrm{crit}$, marking the onset of a continuous parametric transition of non-rotating spherical stars to toroidal configurations, the parameter triple $(h_\mathrm{max}, r_\mathrm{p}/r_\mathrm{e}, \tilde{\beta})$ proves to be particularly useful.
At fixed value $h_\mathrm{max}$, the parameter $\tilde{A}$ can be interpreted as a function of the two variables $r_\mathrm{p}/r_\mathrm{e}$ and $\tilde{\beta}$. As will be discussed below, this function possesses a saddle point with the value  $\tilde{A}=\tilde{A}_\mathrm{crit}$. Hence it can be found by solving the equations
\be\label{e:Det_A_crit}
        \left(\frac{\pt \tilde{A}}{\pt (r_\mathrm{p}/r_\mathrm{e})}\right)_{h_\mathrm{max}} = 0 
        = \left(\frac{\pt \tilde{A}}{\pt \tilde{\beta}}\right)_{h_\mathrm{max}}
\ee
for the corresponding parameters $(r_\mathrm{p}/r_\mathrm{e})_\mathrm{crit}$ and $\tilde{\beta}_\mathrm{crit}$. The associated $\tilde{A}$ assumed for the parameter triple $(h_\mathrm{max}, (r_\mathrm{p}/r_\mathrm{e})_\mathrm{crit}, \tilde{\beta}_\mathrm{crit})$ is the critical value $\tilde{A}_\mathrm{crit}$ in question. In order to determine $\tilde{A}_\mathrm{crit}$ accurately, we compute the function $\tilde{A}=\tilde{A}(r_\mathrm{p}/r_\mathrm{e},\tilde{\beta})$ in a vicinity of the critical point and perform a numerical differentiation with respect to the two variables. We then solve the system (\ref{e:Det_A_crit}) by means of a Newton-Raphson scheme. Note that again a pseudo-spectral treatment ensures a high accuracy of the value $\tilde{A}_\mathrm{crit}$ obtained in this manner.

\subsection{Polytropic bodies}

\begin{figure}
        \includegraphics[scale=1]{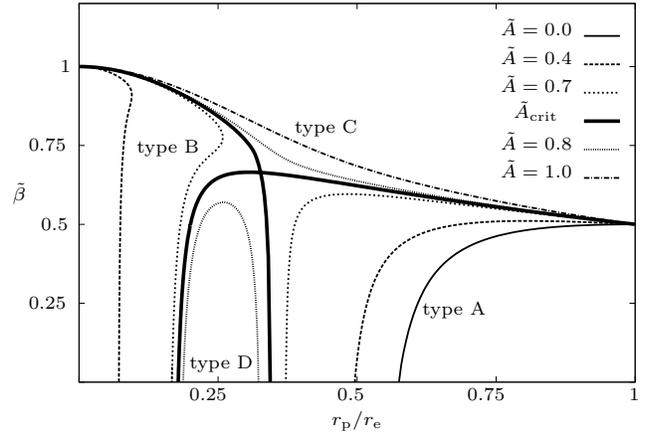}
        \caption{Type A-, B-, C- and D-sequences of differentially rotating polytropic 
				fluid bodies (polytropic index $N=1$) with 
                prescribed maximal enthalpy $\log h_\mathrm{max}=0.2$. Shown is the dependency between the shed parameter $\tilde{\beta}$ and the ratio $r_\mathrm{p}/r_\mathrm{e}$ of polar to equatorial coordinate radius for several values of the rotation law parameter $\tilde{A}$. The bold curve corresponds to the separatrix sequence with $\tilde{A}=\tilde{A}_\mathrm{crit}=0.75904$.}
        \label{fig2}
\end{figure}

In order to illustrate the evolution of the solution space as the rate $\tilde{A}$ of differential rotation is increased, we consider at first fairly weak relativistic polytropic solutions (with polytropic index $N=1$) with a fixed maximal enthalpy, $\log h_\mathrm{max}=0.2$.

For rigid rotation, i.e. $\tilde{A}=0$, the sequence starting from static and
spherical bodies with $r_\mathrm{p}/r_\mathrm{e}=1$ runs through the
spheroidal class and terminates in a mass-shedding limit characterized by
$\tilde{\beta}=0$, see fig. \ref{fig2}. We assign type A to sequences of this
kind and note that the spheroidal class is made up of such sequences.

Associated with the sequence of type A with $\tilde{A}=0$, there is a sequence that contains only configurations with toroidal shape and belongs to the toroidal class. It runs within class II in figure 3.29 of Meinel {\it et al.} \cite{Meineletal08} from the mass-shed curve to the extreme Kerr limit. Since in fig. \ref{fig2} only sequences with spheroidal topology (i.e.~$r_\mathrm{p}>0$) are displayed, this sequence does not appear here.

However, as we move to $\tilde{A}=0.4$ we now find this sequence in
the left part of the figure. It is clearly separated from the
corresponding sequence of type A which runs through the spheroidal
class (in the right part of the figure). This means that, for $\tilde{A}=0.4$, the sequence
belonging to the toroidal class stretches into the regime of bodies
with spheroidal topology, i.e.~the toroidal class contains
configurations with spheroidal topology. We assign type B to sequences
of this kind, that appear for $\tilde{A}$ greater than some threshold value
to be studied in a forthcoming article. In the corresponding situation for $\tilde{A}=0.7$, the
two sequences of types A and B still coexist, but they have moved
towards one another. The merging of the two sequences appears at the
critical value $\tilde{A}_\mathrm{crit}= 0.75904$. The further
increase of $\tilde{A}$ leads again to two separate sequences of types
C and D. The type C sequences show a continuous transition from the
regime of spherical (with $r_\mathrm{p}/r_\mathrm{e}=1$) to that of
toroidal bodies (at $r_\mathrm{p}/r_\mathrm{e}=0$). The type D sequences possess
mass-shedding limits (with $\tilde{\beta}=0$) at both ends. As these two
ends tend towards one another as $\tilde{A}$
is increased, this type of sequence exists only for a small range of values
of the parameter $\tilde{A}$. This range has the lower bound $\tilde{A}_\mathrm{crit}$ and another threshold value as an upper bound (to be studied elsewhere). At this upper bound the sequence of type D degenerates
to a single point describing a particular mass-shedding fluid body. By
contrast, the type C sequences exist for all values
$\tilde{A}>\tilde{A}_\mathrm{crit}$.

The qualitative behaviour for $\log h_\mathrm{max}=0.2$ is typical for the entire solution space. In fig. \ref{fig3} we show sequences for $\tilde{A}\in\{0, 0.1, 0.2, \ldots, 1.5\}$ and $\log h_\mathrm{max}\in\{0, 0.2, 0.4, 0.6\}$, $\log h_\mathrm{max}=0$ meaning the Newtonian limit. The lowermost curves on the right part of the diagrams always correspond to rigid rotation $\tilde{A}=0$, while the uppermost ones represent the configurations with the highest rotation parameter considered, $\tilde{A}=1.5$. 
Also shown (in bold) are the separation sequences corresponding to the values $\tilde{A}_\mathrm{crit}$. From fig. \ref{fig2} it becomes apparent that at the separation point the function $\tilde{A}=\tilde{A}(r_\mathrm{p}/r_\mathrm{e},\tilde{\beta})$ has a maximum with respect to the variable $r_\mathrm{p}/r_\mathrm{e}$ and a minimum with respect to the variable $\tilde{\beta}$, i.e. a saddle point at which (\ref{e:Det_A_crit}) holds. Moreover the curves described by $(\ud^2h/\ud\varrho^2)_c=0$ are plot (dotted lines), which indicate that, for configurations to the left of this curve, the maximal enthalpy is assumed outside the coordinate origin. 

Note that for $\log h_\mathrm{max}=0.6$ {\em two} branches of the separatrix sequence (for which $\tilde{A}_\mathrm{crit}=0.77804$) run into the regime of toroidal figures, as opposed to only one such branch for $\log h_\mathrm{max}\in\{0, 0.2, 0.4\}$. As a consequence, the neighbouring sequences with shedding ends (i.e.~in the diagram for $\log h_\mathrm{max}=0.6$, type D curves with $\tilde{A}=0.8$ and $\tilde{A}=0.9$) possess a branch running into the toroidal regime. However, the type D sequence for $\tilde{A}=1$ is still solely formed of figures with spheroidal topology.

The evolution of the parameter $\tilde{A}_\mathrm{crit}$ as a function of $\log h_\mathrm{max}$ is displayed in fig.  \ref{fig4}. From this figure it can be seen that for a sufficiently large rate of differential rotation, $\tilde{A}>\tilde{A}^\mathrm{N=1}_\mathrm{min}=0.71085$, the spheroidal and toroidal classes merge and hence a continuous parametric transition of spherical to toroidal differentially rotating stars is exhibited. The merging of the classes occurs for $\tilde{A}=\tilde{A}^\mathrm{N=1}_\mathrm{min}$ at $\log h_\mathrm{max}=0.3845$.

\subsection{Homogeneous bodies}
\begin{figure*}
        \includegraphics[scale=1]{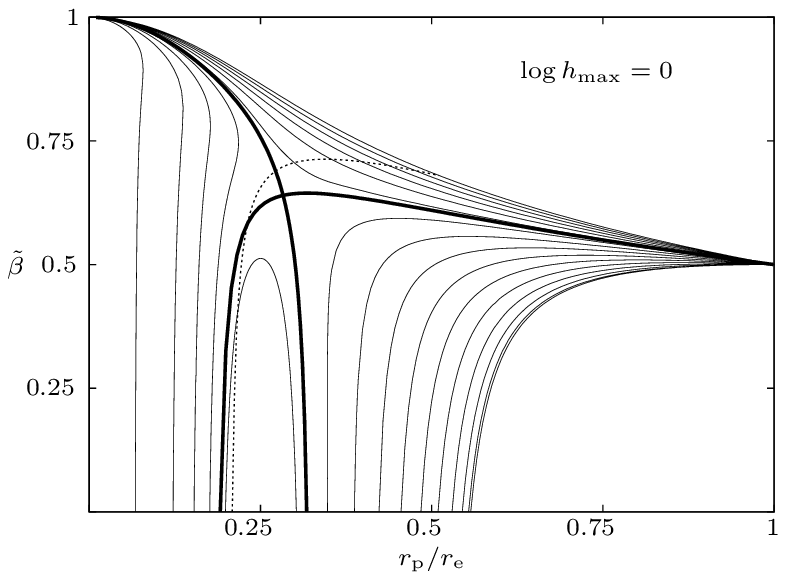}\hfill
        \includegraphics[scale=1]{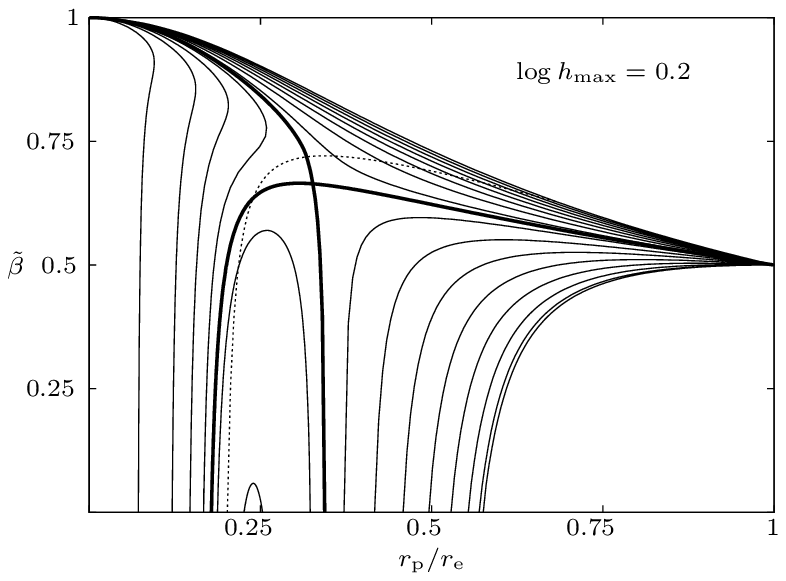}\vspace*{5mm}
        \includegraphics[scale=1]{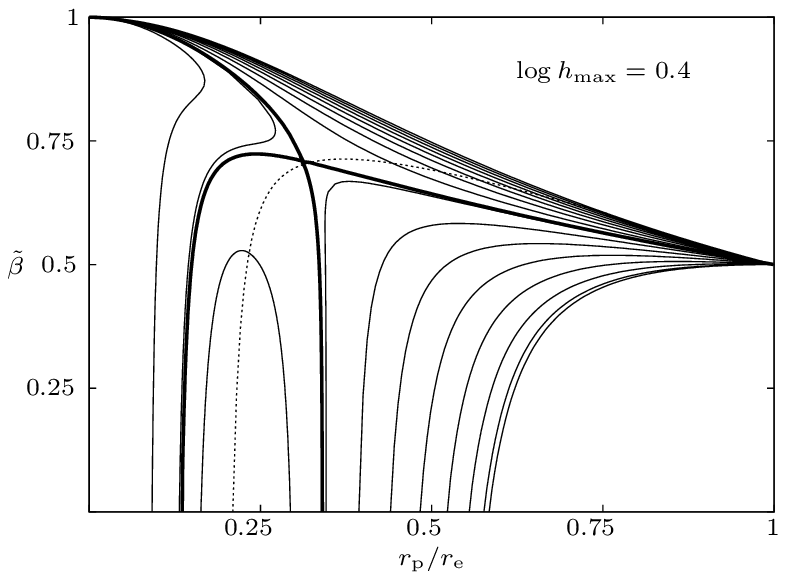}\hfill
        \includegraphics[scale=1]{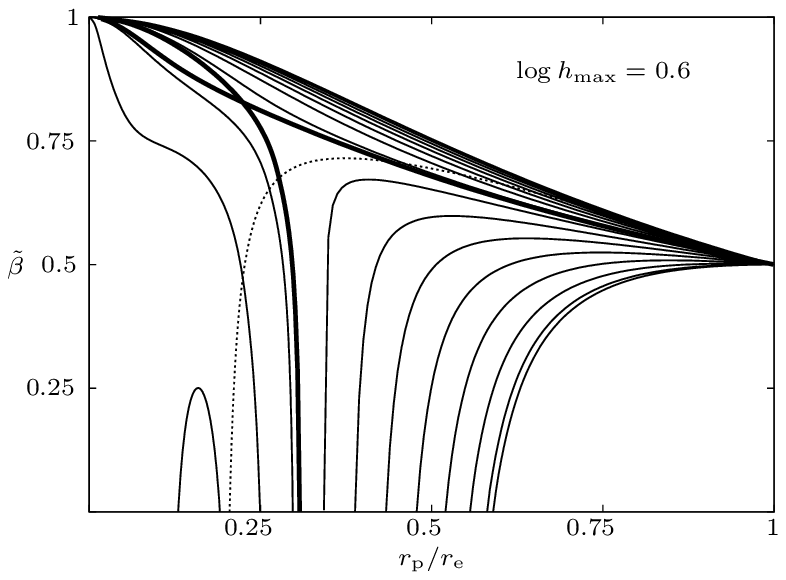}

        \caption{Sequences of differentially rotating polytropic fluid bodies (polytropic index $N=1$) with 
                prescribed maximal enthalpies $\log h_\mathrm{max}$ for $\tilde{A}\in\{0, 0.1, 0.2, \ldots, 1.5\}$. Uppermost and lowermost curves on the right part of the diagrams correspond to rotation parameters $\tilde{A}=1.5$ and $\tilde{A}=0$ (rigid rotation) respectively. The solid  bold curves represent the separatrix sequences with $\tilde{A}=\tilde{A}_\mathrm{crit}$, and the dotted curves defined by $(\ud^2h/\ud\varrho^2)_c=0$ indicate that, for configurations to the left of these curves, the maximal enthalpy is assumed outside the coordinate origin. Note that the diagram for $\log h_\mathrm{max}=0$ displays Newtonian sequences.}
        \label{fig3}
\end{figure*}

For homogeneous and rigidly rotating fluid bodies in General Relativity, the Newtonian limiting sequences have been found to be the separatrix sequences that separate the general-relativistic solution classes from one another, see Ansorg {\it et al.} \cite{Ansorgetal04}, figs 4 and 5 therein. Specifically, the Maclaurin spheroids with $r_\mathrm{p}/r_\mathrm{e}\in[0.17126,1]$ form one branch of the separatrix dividing the spheroidal from the toroidal class. Other branches of this particular separatrix are formed by the Maclaurin spheroids with $r_\mathrm{p}/r_\mathrm{e}\in[0.11160,0.17126]$ and the (non-relativistic) Dyson ring sequence (marked by $A_1^\pm$ in fig. 4 of Ansorg {\it et al.} \cite{Ansorgetal04}).

In the diagram for $\log h_\mathrm{max}=0$, fig.~\ref{fig5}, we find the {\em same} sequences to be the branches of the separatrix which divides particular sequences of differentially rotating Newtonian bodies from one another. We mark this separatrix by a bold curve. The Maclaurin spheroids are characterized by $\tilde{\beta}=\frac{1}{2}$, and the two branches discussed above can easily be identified. The Dyson ring sequence leads from a mass-shedding end $\tilde{\beta}=0$ across the Maclaurin sequence to the regime of toroidal bodies (at $r_\mathrm{p}/r_\mathrm{e}=0$). The remaining piece of the separatrix is given by the sequence $A_2^+$ in fig. 4 of Ansorg {\it et al.} \cite{AKM03b} and leads also from the mass-shedding limit $\tilde{\beta}=0$ to the Maclaurin sequence. 

The separatrix composed of these segments looks qualitatively similar to its polytropic counterpart in fig.~\ref{fig3}. For differential rotation $\tilde{A}>0$ we find sequences of type C in the upper right and of type D in the lower left part of the 
panel, but neither type A nor type B. Note that the only type D curve displayed `inside' the lower separatrix branches corresponds to $\tilde{A}=0.1$, just as the type C curve located opposite and next to the separation crossing point $(r_\mathrm{p}/r_\mathrm{e}=0.17126, \tilde{\beta}=\frac{1}{2})$.

The curves printed in chain lines belong to the Maclaurin sequence with $r_\mathrm{p}/r_\mathrm{e}<0.11160, \tilde{\beta}=\frac{1}{2}$ as well as to the sequence $A_2^-$ in fig. 4 of Ansorg {\it et al.} \cite{AKM03b}, i.e. to branches of separatrices dividing `higher' solution classes from one another. 

In the remaining diagrams of fig.~\ref{fig5} we display sequences for $\tilde{A}\in\{0, 0.1, 0.2, \ldots, 1.5\}$ and $\log h_\mathrm{max}\in\{0.2, 0.4, 0.6\}$. As in fig.~\ref{fig3}, the lowermost curves on the right part of the diagrams always correspond to rigid rotation $\tilde{A}=0$, while the uppermost ones represent the configurations with the highest rotation parameter considered, $\tilde{A}=1.5$. Again, bold lines describe the separatrix sequences, and the dotted lines in all diagrams of fig.~\ref{fig5} are sequences with $(\ud^2h/\ud\varrho^2)_c=0$.

For a given value $\log h_\mathrm{max}$ the qualitative picture for homogeneous bodies is similar to that for polytropes, i.~e.~we can identify sequences of types A and B for $\tilde{A}<\tilde{A}_\mathrm{crit}$ and those of types C and D for $\tilde{A}>\tilde{A}_\mathrm{crit}$. However, the onset of the spheroidal-toroidal transition as a function of $\log h_\mathrm{max}$ is very different, see fig.~\ref{fig6}. If one restricts oneself to differential rotation parameters $\tilde{A}<\tilde{A}^\mathrm{N=1}_\mathrm{min}$, then {\em any} sequence of polytropic bodies starting at the spherical limit $r_\mathrm{p}/r_\mathrm{e}=1$ ends in a mass-shedding limit $\tilde{\beta}=0$ with some value $r_\mathrm{p}/r_\mathrm{e}>0$ (and consequently is of type A). This means that in this range of differential rotation we have a clear gap between the spheroidal and the toroidal class. The distinction between the classes only disappears for $\tilde{A}>\tilde{A}^\mathrm{N=1}_\mathrm{min}$ in which case all four types of sequences can be found. In contrast, the solution space of differentially rotating homogeneous bodies with $\tilde{A}>0$ does not allow at all a classification with respect to spheroidal and toroidal classes. All four types of sequences can be identified in the small range $0<\tilde{A}<\tilde{A}^\mathrm{hom}_\mathrm{max}=0.29646$, and for $\tilde{A}>\tilde{A}^\mathrm{hom}_\mathrm{max}$ we find only type C and D sequences. That is to say that then {\em any} sequence starting at the spherical limit $r_\mathrm{p}/r_\mathrm{e}=1$ leads to the transition to the toroidal regime (and hence is of type C). 
\begin{figure}
        \includegraphics[scale=1]{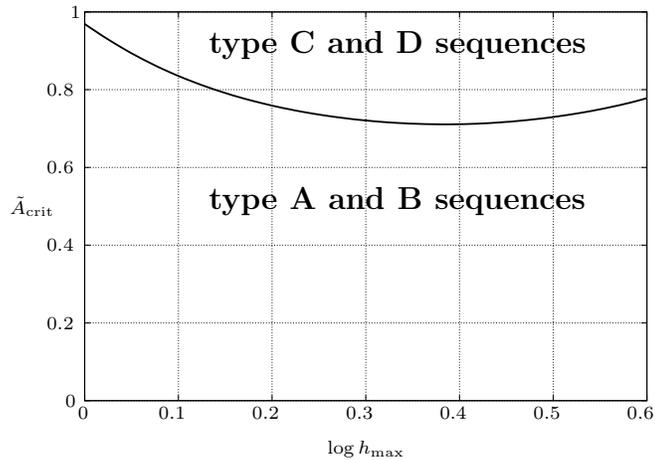}
        \caption{For polytropic bodies (polytropic index $N=1$), the critical rotation parameter $\tilde{A}_\mathrm{crit}$ is plot against the maximal enthalpy, $\log h_\mathrm{max}$. For parameters below this curve, each sequence starting at the spherical and static limit terminates at a mass-shedding limit $\tilde{\beta}=0$. On the other hand, such sequences lead to the transition to the toroidal regime if the parameters chosen are located above the curve.}
        \label{fig4}
\end{figure}
\begin{figure*}
        \includegraphics[scale=1]{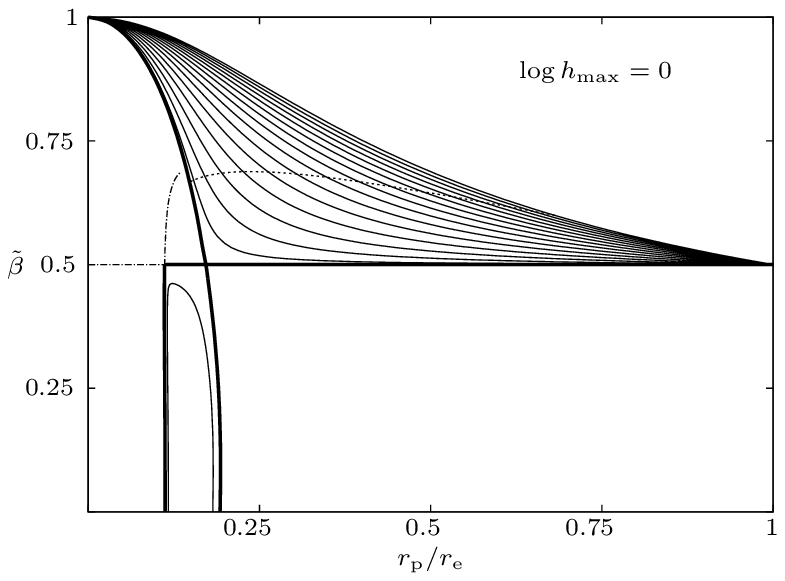}\hfill
        \includegraphics[scale=1]{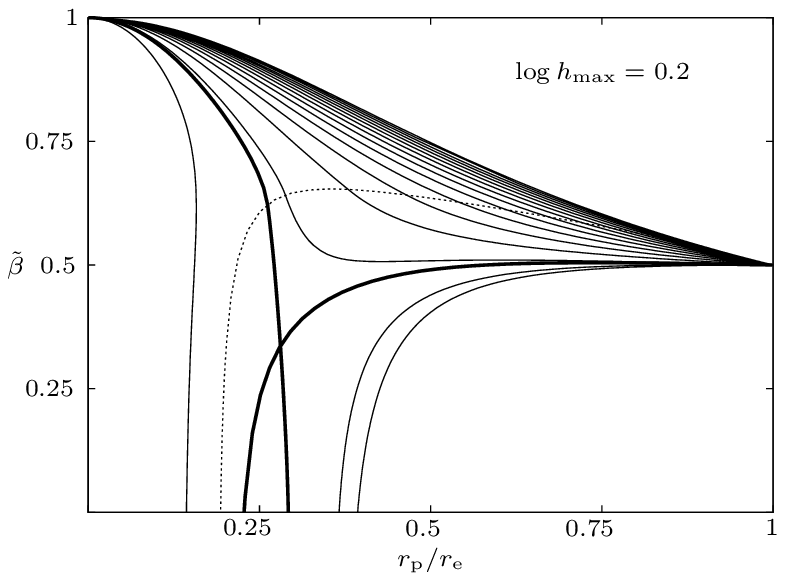}\vspace*{5mm}
        \includegraphics[scale=1]{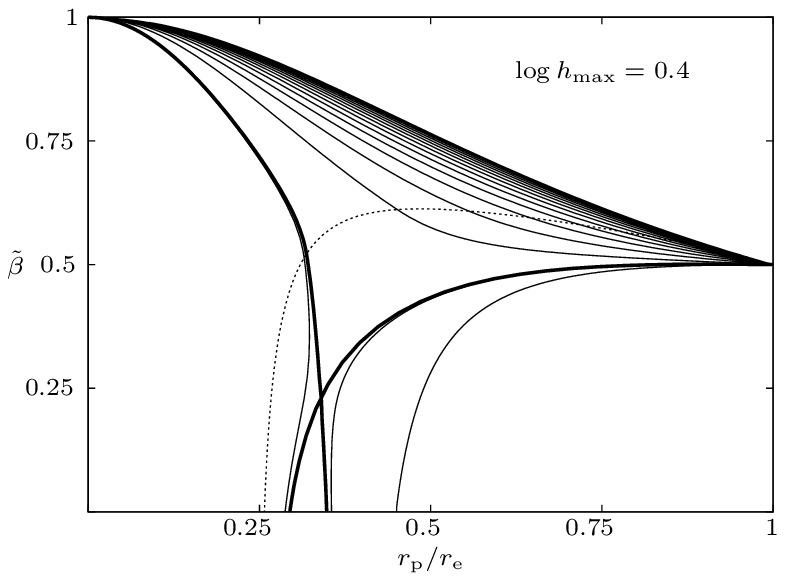}\hfill
        \includegraphics[scale=1]{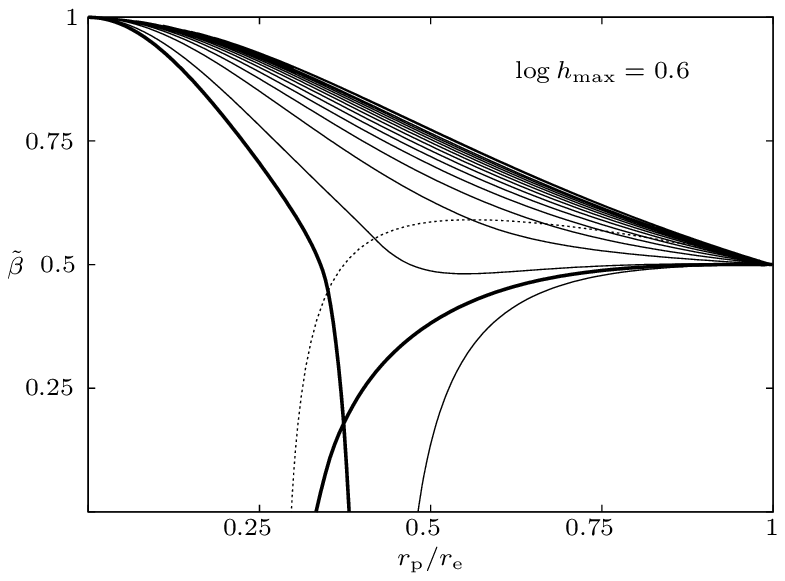}
        \caption{Sequences of differentially rotating homogeneous fluid bodies with 
                prescribed maximal enthalpies $\log h_\mathrm{max}$ for
                $\tilde{A}\in\{0, 0.1, 0.2, \ldots, 1.5\}$. Uppermost and
                lowermost curves on the right part of the diagrams
                correspond to rotation parameters $\tilde{A}=1.5$ and
                $\tilde{A}=0$ (rigid rotation) respectively. As in
                fig.~\ref{fig3}, the solid bold curves represent the
                separatrix sequences with $\tilde{A}=\tilde{A}_\mathrm{crit}$
                and the dotted curves are defined by
                $(\ud^2h/\ud\varrho^2)_c=0$. Note that the diagram for $\log
                h_\mathrm{max}=0$ displays Newtonian sequences, some of which
                divide 'higher' solution classes from one another. See the
                text for more detail.}
        \label{fig5}
\end{figure*}

\begin{figure}
        \includegraphics[scale=1]{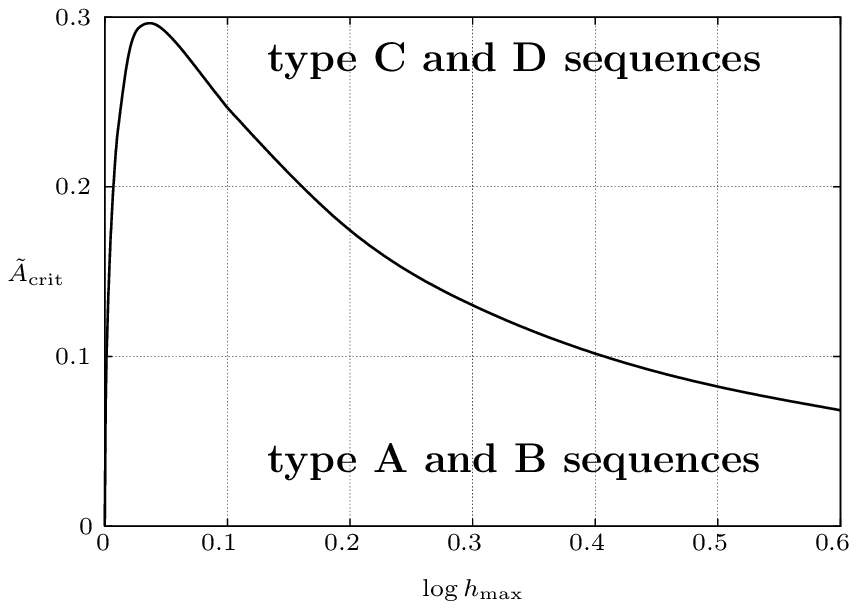}
        \caption{For homogeneous bodies, the critical rotation parameter $\tilde{A}_\mathrm{crit}$ is plot against the maximal enthalpy, $\log h_\mathrm{max}$. As in fig.~\ref{fig4}, parameters below or above this curve describe sequences starting at the spherical and static limit and leading to a mass-shedding limit or to the toroidal regime respectively.}
        \label{fig6}
\end{figure}

\section{Summary and discussion}

In this paper, we have constructed relativistic models of
differentially rotating neutron stars with spheroidal topology (no
hole) for broad ranges of degree of differential rotation
$\tilde{A}\in\{0, 0.1, 0.2, \ldots, 1.5\}$ and for maximal enthalpies
$\log h_\mathrm{max}\in\{0, 0.2, 0.4, 0.6\}$. We have considered homogeneous and 
polytropic configurations (with polytropic index $N=1$) and adopted the rotation law introduced by Komatsu {\it et al.} \cite{KEH}.

From the results obtained we may conclude that the
structure of {\em disjoint classes} in the solution space of
relativistic configurations is a special feature of the sub-space
corresponding to rigid rotation. It turned out that this concept does not survive in the entire solution space
of differentially rotating stars. Instead we find that from a critical rate of differential rotation on, a classification cannot be introduced. For differential rotation larger than this critical rate, the previously disjoint classes have merged and we find continuous transitions from non-rotating and spherical to toroidal objects. 

The critical rotation rate depends strongly on the equation of state being chosen. While the space of sufficiently weakly differentially rotating polytropes still permits the class structure in question, there is no non-vanishing rate of differential rotation for which such a classification makes sense in the case of homogeneous bodies.

We have described properties of the general structure of the solution space in terms of four types of one-dimensional parameter sequences:
\begin{itemize}
	\item 
		Sequences of type A exist for $\tilde{A}< \tilde{A}_\mathrm{crit}$ and consist solely of spheroidal configurations. These sequences start at a static and spherical body and end at the mass shedding limit. Most of these configurations have the maximum enthalpy located in the
		stellar center, but not all of them (for exceptions see e.g.~fig. \ref{fig3}, panel for $\log h_\mathrm{max}=0.6$, type A sequences in the vicinity of the separatrix curve).
	\item 
		Sequences of type B also exist for $\tilde{A}<\tilde{A}_\mathrm{crit}$. They start at the mass
		shedding limit and have a continuous parametric transition to a
		regime of bodies with toroidal topology.  The branch with spheroidal
		topology configurations exists only for $\tilde{A}$ larger than a
		threshold value.
	\item 
		Sequences of type C exist for $\tilde{A}>\tilde{A}_\mathrm{crit}$. They start at a static and spherical body and possess a transition to arbitrarily thin rings. Consequently, type C sequences contain configurations  with spheroidal and toroidal topology.
	\item  
		Sequences of type D also exist for $\tilde{A}>\tilde{A}_\mathrm{crit}$, but only in a narrow range of values of $\tilde{A}$. Their two extremities are configurations at the mass shedding limit.
\end{itemize}
In the solution space, the type A- and B-sequences are clearly separated from the type C- and D-sequences by specific separatrix curves which we determined in a rigorous numerical procedure. These separatrix curves, emphasized through bold solid lines in the figures \ref{fig2}, \ref{fig3} and \ref{fig5}, have in most cases considered three branches that end in the regime of spheroidal bodies and one branch that leads to toroidal bodies. Note that in fig. \ref{fig3}, in the diagram for $\log h_\mathrm{max}=0.6$, an exception can be seen, namely {\em two} branches leading to toroidal topology.

The fact that the classification in classes fails in the realm of differentially rotating bodies means that a new interpretation of the criterion of the maximal mass for the distinction between a black hole and a rotating neutron
star (see e.g.~reasoning in Baumgarte {\it et al.} \cite{Bau00}) becomes necessary. While Baumgarte {\it et al.} \cite{Bau00} discussed an increase of the maximum mass of differentially rotating stars, we find here that the
family of sufficiently strongly differentially rotating bodies does
not possess a maximum at all for the stars' (baryon and gravitational)
masses. Evidently, one can always define one maximal value, restricting the calculation to simply connected bodies, as we did
in this study, but that is not physically relevant since as the full family contains
arbitrarily thin toroidal figures, the mass values may become
arbitrarily large (see Ansorg {\em et al.} \cite{AKM03c} and Fischer
{\em et al.} \cite{Fischeretal05}). In a forthcoming article, we shall come back to this mass criterion and its application to differentially rotating fluids.

In this article the complexity of the full solution space of differentially rotating fluid bodies was demonstrated, which could only be revealed by means of a precise and flexible code that can provide predictions which are relevant from the astrophysical point of view. Indeed, it should be noticed that the critical values obtained for $\tilde{A}$ ($\sim 0.7$) are associated with a degree of differential rotation that is quite similar to what was obtained in dynamical simulations of merging binary neutron stars based on such EOSs (see Shibata \& Uryu \cite{SU}). This means that for the study of properties of stationary configurations, considered as mimics of a specific phase before the collapse to a black hole, the existence of multiple possible configurations has to be kept in mind during the numerical calculations, in order to determine properly an actual sequence and not one built from configurations belonging to different types.

\section*{Acknowledgements}

  This work was supported by the grant SFB/Transregio 7 `Gravitational
  Wave Astronomy' funded by the German Research Foundation; the
  EGO-DIR-102-2007; the FOCUS Programme of Foundation for Polish
  Science, the Polish Grant 1P03D00530; the Polish Astroparticle
  Network 621/E-78/SN-0068/2007; the Associated European
  Laboratory ``Astrophysics Poland-France'' and by the French ANR grant 06-2-134423 ``M\'ethodes math\'ematiques pour la relativit\'e g\'en\'erale''.  LV benefited from the
  Marie Curie Intra-European Fellowship MEIF-CT-2005-025498, within
  the 6th European Community Framework Program.

\newpage

\end{document}